\documentclass[11pt]{article}

\usepackage{amsmath}
\usepackage{graphicx}
\usepackage{enumerate}
\usepackage[numbers, sort, compress]{natbib}
\usepackage{graphics}
\usepackage{algorithm}
\usepackage{algorithmic}
\usepackage{multirow}
\usepackage{subfigure}
\usepackage{verbatim}
\usepackage{listings}
\usepackage{algorithm}
\usepackage{algorithmic}

\title{Simulating the Continuation of a Time Series in R}

\begin{document}

\maketitle

\begin{center}

\large Halis Sak \footnote[1]{Corresponding author. Tel: +90.216.5780000-3363 \\ 
\textit{Email addresses:} halis.sak@gmail.com (Halis Sak), hormannw@boun.edu.tr (Wolfgang H{\"o}rmann)}\\ \normalsize
Department of Industrial and Systems Engineering, Yeditepe University, Kay{\i}\c{s}da\u{g}{\i}, 34755 \.{I}stanbul, Turkey
\\[6pt]

\large Wolfgang H{\"o}rmann\\ \normalsize
Department of Industrial Engineering,
Bo\u{g}azi\c{c}i University,
34342 Bebek-\.{I}stanbul, Turkey
\\[6pt]
\end{center}

\begin{abstract}
  The simulation of the continuation of a given time series is useful for many
  practical applications. But no standard procedure for this task is suggested
  in the literature. It is therefore demonstrated how to use the seasonal ARIMA process to
  simulate the continuation of an observed time series. The R-code presented
  uses well-known modeling procedures for ARIMA models and conditional simulation
  of a SARIMA model with known parameters. A small example demonstrates the correctness
  and practical relevance of the new idea.
\end{abstract}

\noindent%
{\it seasonal ARIMA, simulation, R}

\section{Introduction}

In many application areas it is of practical interest to be able
to simulate the possible continuation of a given time series. For example in
finance the simulation of future stock prices is a well-known standard method
used for risk assessment and for option pricing. 
In inventory management the simulation of future demands could be used to compare the
performance of different inventory policies. Clearly many other examples
of applications in different areas are possible. 
When we wanted to quantify the risk of a company due to the uncertainty of the demand in the next months we
tried to find code that simulates random future observations of the demand subject to
the SARIMA (Seasonal Autoregressive Integrated Moving Average) model we had fitted to the data.
We were astonished when we realized that we were not able to find a single paper in the literature
that tackles this problem.
It is clear that such a simulation conditional on given data requires a modeling and
a parameter estimation step. These two steps are also required for forecasting.
Seasonal (and non-seasonal) ARIMA models have been considered
as standard procedures for many years (\cite{Box;Jenkins:1976a}) and are described in many text books
(see eg. \cite{Montgomery:Jennings:Kulahci;2008a}). After selecting
an ARIMA model and the estimation of its parameters only the conditional simulation
of future realizations given the observations is required.
Many software packages (including R (\cite{R2011})) contain functions
to simulate realizations of ARIMA processes. But 
we were not able to find any description or implementation of
a ``simulation conditional on the observed values" for ARIMA models in the literature.

We therefore present our simple idea of conditional simulating
an SARIMA process in Section~\ref{Sec:idea}. Section~\ref{Sec:Implementation} contains our 
R-implementation whereas Section~\ref{Sec:NumTest} demonstrates the application of our 
code for a practical example.

\section{SARIMA processes}
\label{Sec:idea}

In R, the notation used for ARMA($p,q$) processes is
\begin{equation}
X_{t} = \phi_{1}\, X_{t-1}+...+\phi_{p}\, X_{t-p}+\varepsilon_{t}+\theta_{1}\, \varepsilon_{t-1}+...
+\theta_{q} \,\varepsilon_{t-q} + \mu
\label{model}
\end{equation}
where $\phi_i$ denotes the parameters of the autoregressive process, 
$\theta_j$ the parameters of the moving average process and $\varepsilon$ the white noise error terms
with standard deviation $\sigma$ following the normal distribution.
Using the well known backshift operator notation we can rewrite the above definition by
\[ \phi(B)\, X_t = \theta(B) \, \varepsilon_{t} + \mu\, ,
\]
where $\phi(B)$ denotes a backshift polynomial of order $p$ and $\theta(B)$ a backshift polynomial
of order $q$.
An ARIMA($p,d,q$) process $X_t$ is a process whose $d$-th difference
\[ \nabla ^d \, X_t = (1-B)^d \, X_t
\]  
is an ARMA($p,q$) process. An ARIMA($p,d,q$) process is thus defined by the equation:
\[ \phi(B)\,\nabla^d  X_t = \theta(B) \, \varepsilon_{t}+ \mu \, .
\]
For seasonal ARIMA (SARIMA) processes with period $s$, a seasonal AR polynomial $\Phi(B^s)$ of order $\tilde{p}$, a seasonal MA polynomial $\Theta{B^s}$ of order $\tilde{q}$ and the seasonal difference operator of order $\tilde{d}$
\[ \nabla_s^{\tilde{d}}\, X_t=(1-B^s)^{\tilde{d}} \, X_t 
\]
are required. The SARIMA($p,d,q$)($\tilde{p},\tilde{d},\tilde{q})_s$ is then defined by the equation:

\begin{equation}
\Phi(B^s)\,\phi(B)\,\nabla^{d}\,\nabla_s^{\tilde{d}} X_t = \Theta(B^s) \,\theta(B) \, \varepsilon_{t} + \mu\, .
\label{modelSARIMA}
\end{equation}

For given observations the selection of the model orders $p,d,q,\tilde{p},\tilde{d},\tilde{q}$ for a SARIMA model is the topic of
many books on time series analysis and without the scope of this note. The estimation of the parameters is easy using the
arima() function of the R-base stats package.
We now assume that the observed time series $x_1,x_2,\dots,x_t$ is a realization of the stochastic process defined by all model
assumptions of the SARIMA model and its estimated parameters. We now consider the next $m$
observations of the time series $X=(X_{t+1},X_{t+2},\dots,X_{t+m})$ that are not known yet. The distribution of that
random vector conditional on the observed values $x_1,x_2,\dots,x_t$ is multi-normal and can be called conditional distribution of the future observation. The forecasts of the SARIMA model for the given time series are the expectations of the one- dimensional marginals of that conditional distribution and we write for example
\[ \hat{X}_{t+2} = E(X_t+2|x_1,x_2,\dots,x_t) \, .
\]
The conditional standard deviations of the one-dimensional marginals are used to calculate the prediction error. Exactly these conditional expectations and variances are calculated by the predict() function.

\section{Conditional simulation of a SARIMA process}
\label{Sec:Implementation}

The new idea of this note is now the suggestion to provide code that generates random realizations of the future observation vector conditional on the observed observations. We can write
\[X|x_1,x_2,\dots,x_t =  (X_{t+1},X_{t+2},\dots,X_{t+m})|x_1,x_2,\dots,x_t
\]
for that future observation vector and we hope that the presentation above made clear, why we can call a realization of that random vector   
a random continuation of the time series data we have observed. As the distribution of the vector is multi-normal it would be possible to generate from that distribution calculating its mean vector and variance-covariance matrix. But it is much easier to use directly
the recursion of the model equation of the ARMA model: 
To generate a random realization of $X_{t+1}$ conditional on the past
we need the past observations, the estimated parameters and the residuals (ie. the
estimates of the random shocks $\varepsilon_i$). It is then no problem to simulate
$X_{t+1}$ using the recursion given in formula (\ref{model}). The new random shock $\varepsilon_{t+1}$ 
is generated as a normal random variate with mean zero and standard deviation $\sigma$. 
The simulation of $X_{t+2}$ is done conditional on the past observations and on the generated
values $\varepsilon_{t+1}$ and $X_{t+1}$. 

For the case of SARIMA processes the model equation is again a linear combination of past observations and past random shocks together with the new random shock $\varepsilon_t$. Due to the seasonal model some of the AR-parameters ($\phi$) and MA-parameters
($\theta$) are equal to zero.  So again we can use the recursive approach explained above.

\medskip

In this code snippet we present the R routine we coded according to the above explanations.
It generates future observations from seasonal and non-seasonal ARIMA processes conditional on an observed time series.

Algorithm~\ref{Alg:Sim} summarizes how $m$ future values are simulated from seasonal and 
non-seasonal ARIMA process. When we fit the ARIMA model using the  
arima() function of the R-base stats package, all the model parameters including
the estimated variance of error terms ($\sigma^2)$ are returned as 
demonstrated in Section~\ref{Sec:NumTest}. R codes for Algorithm~\ref{Alg:Sim} are given in Section~\ref{SecSourceCode}.

\begin{algorithm} 
\caption{simulating $m$ future observations from seasonal and non-seasonal ARIMA process.} 
\label{Alg:Sim} 
\begin{algorithmic}[1]
\STATE If necessary do the differencing (seasonal and/or non-seasonal) of the data given in the fitted model (otherwise
$\nabla x$ simply refers to the original series)
\STATE compute intercept $\mu = \bar{\nabla x}(1-\sum_{i=1}^p \phi_{i})$ where $\bar{\nabla x}$ denotes for
the average of $\nabla x$
\STATE construct a vector $x$ of size $p+m$
\STATE construct a vector $\varepsilon$ of size $q+m$
\STATE equate first $p$ terms of $x$ to $\nabla x$ at newest $p$ time steps 
\STATE equate ($\varepsilon_{[1]},...,\varepsilon_{[q]}$) to the residuals of data at newest $q$ time steps 
\FOR{future time steps $k=1,..,m$} 
\STATE generate $\varepsilon_{[q+k]}$ from $N(0, \sigma)$
\STATE apply moving average and auto-regressive filtering on $x_{[p+k]}$ as in (Equation~\ref{model}) 
\ENDFOR
\STATE remove first $p$ elements of $x$ to get only differences of future time steps
\STATE undifference x
\STATE return x
\end{algorithmic}
\end{algorithm}

\section{Numerical experiments}
\label{Sec:NumTest}

In this section we first fit a seasonal model to monthly totals of 
international airline passengers between $1949$ and $1960$ using the arima(). (The data are available in the R 
package fma \cite{fma:2009}.)

\begin{verbatim}
R> library("fma")
R> set.seed(4321)
R> data <- airpass
R> Par <- c(1, 1, 1, 0, 1, 0)
R> fit <- arima(data, order = c(Par[1], Par[2], Par[3]),
+           seasonal = list(order = c(Par[4], Par[5], Par[6])))
R> fit
Series: data 
ARIMA(1, 1, 1)(0, 1, 0)[12]                    
Coefficients:
          ar1      ma1
      -0.3009  -0.0073
s.e.   0.3835   0.4133
sigma^2 estimated as 137:  log likelihood = -508.2
AIC = 1022.39   AICc = 1022.58   BIC = 1031.02
\end{verbatim}

\begin{figure}[htbp]
\begin{center}
\includegraphics*[scale = 0.45]{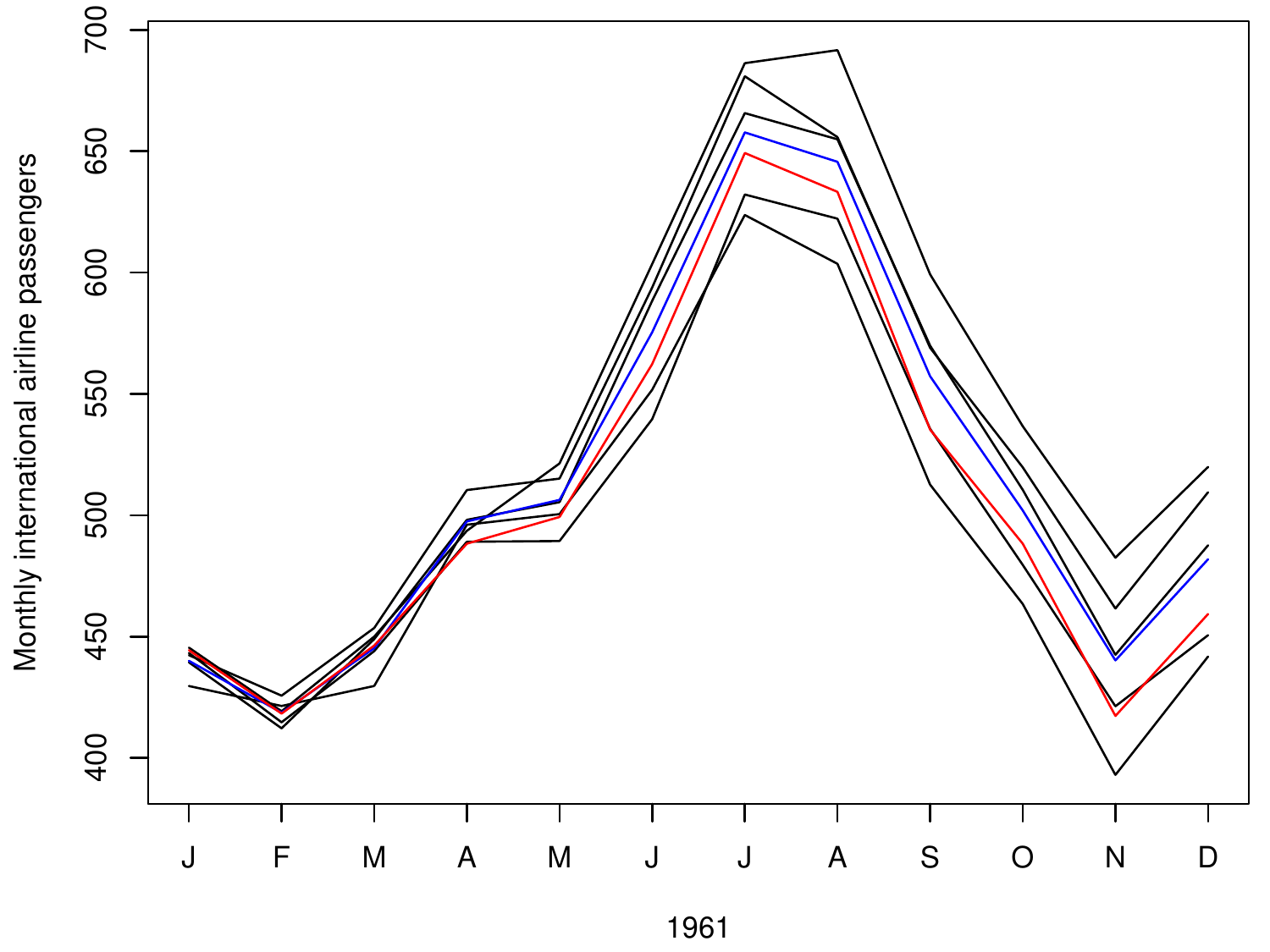}
\caption{Five different simulations and their average (blue line) of the monthly international airline passengers in $1961$ and forecasted values (red line).}
\label{FiveSims}
\end{center}
\end{figure}

For demonstration purposes we generate five different independent continuations of the 
time series and show them, their average and the forecasted values in Figure~\ref{FiveSims}. 

\begin{verbatim}
R> sims <- arima.condsim(fit, data, n.ahead = 12, n = 5)
R> ts1 <- ts(sims[, 1], f = frequency(data), s = tsp(data)[2] + 
+         1/tsp(data)[3])
R> ts2 <- ts(sims[, 2], f = frequency(data), s = tsp(data)[2] + 
+         1/tsp(data)[3])
R> ts3 <- ts(sims[, 3], f = frequency(data), s = tsp(data)[2] + 
+         1/tsp(data)[3])
R> ts4 <- ts(sims[, 4], f = frequency(data), s = tsp(data)[2] + 
+         1/tsp(data)[3])
R> ts5 <- ts(sims[, 5], f = frequency(data), s = tsp(data)[2] + 
+         1/tsp(data)[3])
R> tsA <- ts(sapply(seq_len(12), function(i) mean(sims[i,])), 
+         f = frequency(data), s = tsp(data)[2]+1/tsp(data)[3])
R> ts.plot(ts1, ts2, ts3, ts4, ts5, gpars = list(xlab = "1961", 
+          ylab = "Monthly international airline passengers", xaxt = "n"))
R> lines(tsA, col="blue")
R> lines(predict(fit, n.ahead=12)$pred, col="red")
R> axis(1, time(ts1), rep(substr(month.abb, 1, 1), length = length(ts1)))
\end{verbatim}

In the following experiment we simulate $10,000$ independent continuations and show that, as expected, 
the mean of the simulated values is very close to the forecasted values. It is also possible to use innovations 
equal to zero in our function arima.condsim() to produce the exact 
forecasts. 

\begin{verbatim}
R> sims <- arima.condsim(fit, data, n.ahead = 12, n = 10000)
R> sims_mean <- sapply(seq_len(12), function(i) mean(sims[i, ]))
R> ts <- ts(sims_mean, f = frequency(data), s = tsp(data)[2] + 
+         1/tsp(data)[3])
R> ts

       Jan      Feb      Mar       Apr       May    Jun    
1961 444.2828 418.1049 446.0237 487.9601 498.8899 562.0800

       Jul      Aug      Sep       Oct       Nov    Dec
1961 648.9706 633.0297 535.0563 487.9923 417.1746 459.2555
R> predict(fit, n.ahead = 12)

$pred
       Jan    Feb      Mar       Apr       May     Jun    
1961 444.3670 418.2566 446.2898 488.2798 499.2828 562.2819
       Jul    Aug      Sep       Oct       Nov     Dec
1961 649.2822 633.2821 535.2821 488.2821 417.2821 459.2821
\end{verbatim}

Finally, we fit a non-seasonal model to the same data to show that our function
works both for seasonal and non-seasonal models.

\begin{verbatim}
R> Par <- c(1, 0, 1, 0, 0, 0)
R> fit <- arima(data, order = c(Par[1], Par[2], Par[3]),
+           seasonal = list(order = c(Par[4], Par[5], Par[6])))
R> fit
Series: data 
ARIMA(1, 0, 1) with non-zero mean
Coefficients:
          ar1     ma1     intercept
        0.9373  0.4264    281.5426
s.e.    0.0302  0.0911     53.6135
sigma^2 estimated as 968.5:  log likelihood = -700.87
AIC = 1409.75   AICc = 1410.04   BIC = 1421.63
R> sims <- arima.condsim(fit, data, n.ahead = 12, n = 10000)
R> sims_mean <- sapply(seq_len(12), function(i) mean(sims[i, ]))
R> ts <- ts(sims_mean, f = frequency(data), s = tsp(data)[2] + 
+         1/tsp(data)[3])
R> ts

       Jan      Feb      Mar       Apr       May    Jun    
1961 453.9091 443.5161 432.8683 422.7560 414.1958 406.3113

       Jul      Aug      Sep       Oct       Nov    Dec
1961 398.7037 391.8506 384.9362 378.4532 372.7470 367.1855
R> predict(fit, n.ahead = 12)

$pred
       Jan    Feb      Mar       Apr       May     Jun    
1961 453.9038 443.0989 432.9713 423.4785 414.5809 406.2410
       Jul    Aug      Sep       Oct       Nov     Dec
1961 398.4239 391.0969 384.2292 377.7920 371.7583 366.1029
\end{verbatim}

\section{Source code}
\label{SecSourceCode}
\begin{verbatim}
arima.condsim <- function(object, x, n.ahead = 1, n = 1){
    L <- length(x); coef <- object$coef; 
    arma <- object$arma; model <- object$model; 
    p <- length(model$phi); q <- length(model$theta)	
    d <- arma[6]; s.period <- arma[5]; 
    s.diff <- arma[7]
		
    if(s.diff > 0 && d > 0){
        diff.xi <- 0; 
        dx <- diff(data, lag = s.period, differences=s.diff)
        diff.xi[1] <- dx[length(dx) - d + 1]; 
        dx <- diff(dx, differences = d)
        diff.xi <- c(diff.xi[1], data[(L - s.diff * s.period + 1):L])
    }else if(s.diff > 0){
        dx <- diff(data, lag = s.period, differences = s.diff)	
        diff.xi <- data[(L - s.diff * s.period + 1):L]	
    }else if(d > 0){
        dx <- diff(data, differences = d); 
        diff.xi <- data[(L - d + 1):L]	 			
    }else{dx <- data}

    use.constant <- is.element("intercept", names(coef))
    mu <- 0	
    if(use.constant){
        mu <- coef[sum(arma[1:4]) + 1][[1]] * (1 - sum(model$phi))	
    }
    p.startIndex <- length(dx) - p
    start.innov <- NULL
    if(q > 0){
        start.innov <- residuals(object)[(L - q + 1):(L)]
    }	

    res <- array(0, c(n.ahead, n))

    for(r in 1:n){
    innov = rnorm(n.ahead, sd = sqrt(object$sigma2)) 
    if(q > 0){ 
        e <- c(start.innov, innov)
    }else{e <- innov}

    xc <- array(0, dim = p + n.ahead)
    if(p != 0) for(i in 1:p) xc[i] <- dx[[p.startIndex + i]]

    k <- 1 
    for(i in (p + 1):(p + n.ahead)){
       xc[i] <- e[q + k]			
       if(q != 0)
           xc[i] <- xc[i] + sum(model$theta * e[(q + k - 1):k]) 			
       if(p != 0)
           xc[i] <- xc[i] + sum(model$phi * xc[(i - 1):(i - p)])
       if(use.constant)
           xc[i] <- xc[i] + mu
        k <- k + 1
    }	
    xc <- as.vector(unlist(xc[(p + 1):(p + n.ahead)]))

    if((d > 0) && (s.diff > 0)){
        xc <- diffinv(xc, differences = d, xi = diff.xi[1])[-c(1:d)]
        xc <- diffinv(xc, lag = s.period, differences = s.diff,
                     xi = diff.xi[2:(s.diff * s.period + 1)])
        xc <- xc[-(1:(s.diff * s.period))]		
    }else if(s.diff > 0) {
        xc <- diffinv(xc, lag = s.period, differences = s.diff, 
                       xi = diff.xi[1:(s.diff * s.period)])
        xc <- xc[-(1:(s.diff * s.period))]				
    }else if(d > 0){
        xc <- diffinv(xc, differences = d, xi = diff.xi)[-c(1:d)]
    }    
    res[, r] <- xc
    }
    res
}
\end{verbatim}

\section{Discussion}

We have demonstrated that, using a SARIMA model, it is not difficult to simulate from the conditional distribution of future observations. Our code can thus be used to randomly generate possible future continuations of a time series. The identification of a suitable SARIMA model is an important step in the procedure we suggest; due to the nature of this short note we have to refer the reader to the vast literature on time series analysis for this task; an important point in the modeling procedure are also checks for the model assumption. Especially the normal assumption for the error term has an important impact on the simulated future observations.

Despite these important limitations, that are present in all parametric statistical models, we hope that our simple algorithm will be useful for many applications. This seems likely as we were not able to find any suggestions for a similar algorithm in the literature.

\newpage

\end{document}